\documentclass[aps,groupaddress,floats,preprint,showpacs]{revtex4}
\usepackage{amsfonts}
\usepackage{amssymb}
\usepackage{bm}
\usepackage[final]{graphicx}

\begin{document}
\bibliographystyle{apsrev}


\newcommand{\alps}{\ensuremath{\alpha_s}}
\newcommand{\qbar}{\bar{q}}
\newcommand{\beq}{\begin{equation}}
\newcommand{\eeq}{\end{equation}}
\newcommand{\beqa}{\begin{eqnarray}}
\newcommand{\eeqa}{\end{eqnarray}}
\newcommand{\gs}{g_{\pi NN}}
\newcommand{\gw}{f_\pi}
\newcommand{\mq}{m_Q}
\newcommand{\mn}{m_N}
\newcommand{\bb}{\langle}
\newcommand{\kb}{\rangle}
\newcommand{\st}{\ensuremath{\sqrt{\sigma}}}
\newcommand{\rvec}{\mathbf{r}}
\newcommand{\bvec}[1]{\ensuremath{\mathbf{#1}}}
\newcommand{\bra}[1]{\ensuremath{\bb#1|}}
\newcommand{\ket}[1]{\ensuremath{|#1\kb}}
\newcommand{\gft}{\ensuremath{\gamma_{FT}}}
\newcommand{\bfalp}{\mbox{\boldmath{$\alpha$}}}
\newcommand{\bfnab}{\mbox{\boldmath{$\nabla$}}}
\newcommand{\bfpi}{\mbox{\boldmath{$\pi$}}}
\newcommand{\bfsig}{\mbox{\boldmath{$\sigma$}}}
\newcommand{\bftau}{\mbox{\boldmath{$\tau$}}}
\newcommand{\spup}{\uparrow}
\newcommand{\spd}{\downarrow}
\newcommand{\hbarom}{\frac{\hbar^2}{m_Q}}
\newcommand{\half}{\frac{1}{2}}
\newcommand{\vnn}{\ensuremath{\hat{v}_{NN}}}
\newcommand{\argonne}{\ensuremath{v_{18}}}
\newcommand{\lqcd}{\ensuremath{\mathcal{L}_{QCD}}}
\newcommand{\lgf}{\ensuremath{\mathcal{L}_g}}
\newcommand{\lqm}{\ensuremath{\mathcal{L}_q}}
\newcommand{\lqg}{\ensuremath{\mathcal{L}_{qg}}}
\newcommand{\nn}{\ensuremath{NN}}
\newcommand{\hpnd}{\ensuremath{H_{\pi N\Delta}}}
\newcommand{\hpqq}{\ensuremath{H_{\pi qq}}}
\newcommand{\fpnn}{\ensuremath{f_{\pi NN}}}
\newcommand{\fpnd}{\ensuremath{f_{\pi N\Delta}}}
\newcommand{\fpqq}{\ensuremath{f_{\pi qq}}}
\newcommand{\ylm}{\ensuremath{Y_\ell^m}}
\newcommand{\ylmc}{\ensuremath{Y_\ell^{m*}}}
\newcommand{\qbh}{\hat{\bvec{q}}}
\newcommand{\xbh}{\hat{\bvec{X}}}
\newcommand{\dt}{\Delta\tau}
\newcommand{\qmag}{|\bvec{q}|}
\newcommand{\pmag}{|\bvec{p}|}
\newcommand{\oas}{\ensuremath{\mathcal{O}(\alpha_s)}}
\newcommand{\vtxb}{\ensuremath{\Lambda_\mu(p',p)}}
\newcommand{\vtxp}{\ensuremath{\Lambda^\mu(p',p)}}
\newcommand{\pwqp}{e^{i\bvec{q}\cdot\bvec{r}}}
\newcommand{\pwqm}{e^{-i\bvec{q}\cdot\bvec{r}}}
\newcommand{\gsa}[1]{\ensuremath{\bb#1\kb_0}}
\newcommand{\oer}[1]{\mathcal{O}\left(\frac{1}{\qmag^{#1}}\right)}
\newcommand{\nub}[1]{\overline{\nu^{#1}}}
\newcommand{\balph}{\mbox{\boldmath{$\alpha$}}}
\newcommand{\bgam}{\mbox{\boldmath{$\gamma$}}}
\newcommand{\epf}{E_\bvec{p}}
\newcommand{\epfp}{E_{\bvec{p}'}}
\newcommand{\eka}{E_{\alpha\kappa}}
\newcommand{\ekaq}{(E_{\alpha\kappa})^2}
\newcommand{\ekap}{E_{\alpha'\kappa}}
\newcommand{\ekpa}{E+{\alpha\kappa_+}}
\newcommand{\ekma}{E_{\alpha\kappa_-}}
\newcommand{\ekp}{E_{\kappa_+}}
\newcommand{\ekm}{E_{\kappa_-}}
\newcommand{\ekpap}{E_{\alpha'\kappa_+}}
\newcommand{\ekmap}{E_{\alpha'\kappa_-}}
\newcommand{\yjm}[1]{\mathcal{Y}_{jm}^{#1}}
\newcommand{\ysa}[3]{\mathcal{Y}_{#1,#2}^{#3}}
\newcommand{\ysc}{\tilde{y}}

\title
{Electromagnetic response of confined Dirac particles}

\author{Mark W.\ Paris}
\email[]{paris@lanl.gov}
\affiliation{Theoretical Division,
Los Alamos National Laboratory,
MS B283, Los Alamos, NM 87545}
\affiliation{Departement f\"{u}r Physik und Astronomie,
Universit\"{a}t Basel, Switzerland}

\date{\today}

\begin{abstract}
\medskip
The eigenstates of a single massless Dirac particle confined in a 
linear potential are calculated exactly by direct solution of the
Dirac equation. The electromagnetic structure functions are calculated 
from the Dirac wave functions of the ground and excited states of 
the particle by coupling to its conserved vector current. We obtain 
the longitudinal and transverse structure functions as a function 
of $\ysc=\nu-\qmag$, where $\nu$ and $\qmag$ are the energy and 
momentum transferred to the target in its rest frame. At values of 
$\qmag \gtrsim 2.5$ GeV, 
much larger than the characteristic energy scale $\sim\!\!$ 440 MeV
of the confining potential, the response exhibits $\ysc$ scaling, 
a generalization of 
Bjorken scaling. We compare the exact structure functions with
those obtained from the ground state wave functions in the plane
wave impulse approximation. The deviation from the Callan-Gross 
relation is compared with the parton model prediction.
\end{abstract}
\maketitle

\section{Introduction}
The present work is motivated by the study of the effects of
interactions on the response of systems of confined relativistic
particles. Here we consider a single massless Dirac particle confined
by a linear potential. The Dirac Hamiltonian is solved exactly for a
large number of eigenstates and the electromagnetic structure 
functions are calculated by coupling a charged leptonic probe 
(such as the electron) to the conserved vector current of the 
target, $\bar{\psi}_q(x)\gamma_\mu\psi_q(x)$.
This model neglects some of the
important physics such as the contribution of 
$q\bar{q}$ pairs from the vacuum and the effects of radiative
gluon corrections. It is nevertheless useful to study the role
of interactions on valence quark structure functions.
In a future paper we will use this model to investigate the 
role of interactions in determining spin-dependent observables.\cite{PPS03}

The model may be viewed as a heavy-light meson, such as $\bar{t}u$,
or a baryon with a light quark bound to an infinitely massive diquark.
In the heavy-light meson, the antiquark $\bar{t}$ and light quark $u$
have opposite color charge and confinement is due to the flux tube
connecting them. In the limit of very large mass of the $\bar{t}$ the
Dirac equation for the motion of the light quark is exact. We neglect
the weak interaction.

Benhar, Pandharipande,
and Sick \cite{BPS00} have shown that the transverse structure function 
of the proton $W^p_1(\ysc,\qmag)$, measured in inclusive deep inelastic 
scattering (DIS) experiments, exhibits scaling in the variable
$\ysc=\nu-\qmag$ over a large range, 5 GeV $\le\qmag\le$ 100 GeV. Here
$\nu$ and $\qmag$ are the energy and momentum transferred to the target
in its rest frame. The model described in the present work also 
exhibits $\ysc$ scaling. The $\ysc$ variable is related to the 
dimensionless Nachtmann 
variable \cite{ON} $\xi=-\ysc/M$, where $M$ is the target mass. The 
variable $\ysc$ is appropriate to systems where the constituents of 
the target move relativistically.\cite{Paris02} It and the Nachtmann 
$\xi$ are generalizations \cite{BPS00,Paris01} of the Bjorken scaling
variable $x_B$. Their relation is given by:
\beq
\label{eqn:svrel}
\xi=-\frac{\ysc}{M}=\frac{2x_B}{1+\sqrt{1+\frac{4x_B^2M^2}{Q^2}}},
\eeq
with $Q^2=\qmag^2-\nu^2$, giving $\xi=x_B$ in the limit
$Q^2\rightarrow\infty$.

In Ref.\cite{Paris01} we calculated the exact response of a single 
relativistic scalar ``quark'' confined by a linear potential. The
response of this simple model exhibited a rich behavior and allowed
the study of a diverse range of phenomena. We observed the $\ysc$ 
scaling of the response in the spacelike and timelike regions of 
momentum transfer, an approach to scaling qualitatively similar to
the one found in 
recent inclusive experiments of DIS of electrons by proton
\cite{Keppel}, and Bloom-Gilman duality \cite{Bloom}. We also 
studied various approximations to the exact result. In particular
the on-shell approximation (OSA), which assumes that constituents
of a hadronic target are on the mass-shell before and after 
interaction with the probe, admits response only in the spacelike 
region of momentum transfer due to the inequality
\beq
\label{eqn:osa}
\nu = \sqrt{|\bvec{k}+\bvec{q}|^2 + m^2} -\sqrt{|\bvec{k}|^2 + m^2}
\leqslant \qmag,
\eeq
where $\bvec{k}$ is the momentum of the struck constituent and $m$ is
its mass. In contrast, the plane wave impulse approximation (PWIA), 
often used in nuclear physics, treats the initial state constituents 
as bound particles, which are therefore not on their mass-shell, and
Eq.(\ref{eqn:osa}) is not satisfied. Thus there is response in the 
timelike region in PWIA due to the bound nature of the constituents.
Final state interactions (FSI), however, are neglected in PWIA and
the state of the particle after it is struck by the probe is
described as a plane wave.

In Ref.\cite{Paris02} we investigated the role of FSI 
effects on the response. There we found that FSI broadens 
the PWIA response, shifting more strength into the timelike 
region of momentum transfer. This broadening of the response
due to FSI persists as $\qmag\rightarrow\infty$.
The exact response was shown 
to be calculable by convolution of the PWIA response with a
``folding function.'' This function describes the distribution
in energy of a plane wave state in the linear confining
well. It becomes independent of the three-momentum $\qmag$ 
transferred to the target in the limit of large
$\qmag$ ({\em i.e.} the ``scaling'' limit). Thus the $\ysc$ 
scaling, and equivalently Nachtmann $\xi$
scaling or Bjorken scaling of the
exact response is not a signal that FSI may be neglected.
In fact, in the semi-relativistic model of Ref.\cite{Paris02}
it is crucial to take into account FSI in order to reproduce 
the response {\em even in the scaling limit}, 
$\qmag\rightarrow\infty$. This situation is distinct from the
case of a confining potential with a non-relativistic 
kinetic energy. In such a model, the width of the response 
scales linearly with $\qmag$. This precludes the need to take 
FSI into account in the scaling limit. It is also interesting 
to note that there is no Bloom-Gilman duality in the non-relativistic 
model. Since the width of the response scales linearly
with $\qmag$ the resonant structure of the response at low-$\qmag$
does not average around the more broad response at high-$\qmag$.

The semi-relativistic model leaves much to be desired. Among
its shortcomings are an ill-defined current operator, 
neglect of the contribution of $q\bar{q}$ pairs, and neglect
of radiative gluon effects. In this work
we address the most egregious of these shortcomings by working
with the conserved electromagnetic current of the confined 
particles, $\bar{\psi_q}(x)\gamma_\mu\psi_q(x)$.

We proceed by solving the Dirac equation for an assumed
potential
\beqa
\label{eqn:potdef}
V &=& V(r)\half(1+\beta) =
 V(r) \left(\begin{array}{cc} 1 & \ 0 \\ 0 & \ 0 \end{array}\right) \\
\label{eqn:radpotdef}
V(r) &=& \st r.
\eeqa
This potential has been investigated by Page, Ginocchio, and Goldman
\cite{Page:2000ij} and shown to admit a {\em spin symmetry} which 
suppresses spin-orbit coupling in the hadron spectrum. This 
symmetry results in the degeneracy of states with the same orbital
angular momentum $\ell$ of the upper component of
the Dirac wave function. The states with $j=\ell\pm\half$ are
degenerate and they have Dirac quantum numbers $\kappa=-(\ell+1)$
and $\ell$, respectively.  Recall that $\kappa=\pm(j+\half)$ where
the upper (lower) sign is used when $j=\ell\mp\half$.
This potential is easy to work
with since the lower components of the wave function are not 
coupled by the Dirac matrix of Eq.(\ref{eqn:potdef}).

In the next section we will briefly detail the solution of the
single particle Dirac equation for this potential. Section 
\ref{sec:sf} gives the calculation of the electromagnetic 
structure functions. We discuss the interpretation of the 
contribution to the structure functions of the negative energy 
states. Section \ref{sec:PWIA} briefly compares the exact
structure functions to those obtained in the PWIA. In the
concluding Sec. \ref{sec:concl}, we study the deviation 
from the Callan-Gross relation.

\section{Diagonalization in the static bag model basis}
\label{sec:diag}
We wish to solve the Dirac equation
\beq
\label{eqn:Deq}
H\Psi^P_{n\kappa m}=\left[\bfalp\cdot\bvec{p} + V \right]
\Psi^{P}_{n\kappa m} = E^{P}_{n\kappa} \Psi^{P}_{n\kappa m}
\eeq
for a massless fermion in the potential given 
in Eq.(\ref{eqn:potdef}). Here $n,\kappa$,
and $m$ are the quantum numbers of a Dirac particle in
a spherically symmetric potential. The Eq.(\ref{eqn:Deq}) with
potential given by Eqs.(\ref{eqn:potdef},\ref{eqn:radpotdef})
has a discrete spectrum of bound positive energy $P=+$ states
with wave functions that vanish as $r\rightarrow\infty$.
However, due to the form of the Dirac matrix in Eq.(\ref{eqn:potdef}) 
the wave functions of negative energy states are unbound.

We expand the eigenstates $\Psi^{P}_{n\kappa m}$
in a complete basis of static bag model 
(herein ``bag model'') states $\Phi^{P}_{\alpha\kappa m}$.
(The explicit form of the bag model states $\Phi^{P}_{\alpha\kappa m}$
is given in the Appendix \ref{sec:app1}.) The expansion is
\beq
\label{eqn:exp}
\Psi^{P}_{n\kappa m}
= \sum_{\alpha=1,\ldots,\infty}
[A^{+}_{n\kappa\alpha} \Phi^{+}_{\alpha\kappa m}
+A^{-}_{n\kappa\alpha} \Phi^{-}_{\alpha(-\kappa)m}].
\eeq
Here we have explicitly written out the positive, 
$\Phi^+_{\alpha\kappa m}$ and negative, $\Phi^-_{\alpha(-\kappa)m}$
energy terms in the expansion of the eigenstates
of $H$. The bag model states, $\Phi^{P}_{\alpha\kappa m}$
are eigenstates of $\mathcal{K}=\beta(\bfsig\cdot\bvec{L}+1)$ 
with eigenvalue
\beqa
\label{eqn:KP}
\mathcal{K}\Phi^{P}_{\alpha\kappa m} = - \mbox{sgn}\, P \kappa
\Phi^{P}_{\alpha\kappa m}.
\eeqa
The potential mixes $P=+$ and $P=-$ bag model states with 
opposite signs of $\kappa$.

In the bag model basis the Dirac equation is
\beq
\label{eqn:depb}
\sum_{P',P,\kappa',\kappa,\alpha',\alpha} \left[
\bra{\Phi^{P'}_{\alpha'\kappa' m}} V \ket{\Phi^{P}_{\alpha\kappa m}}
+ ( E^P_{n\kappa} - \mbox{sgn} P E_{\alpha\kappa} )
\delta_{P',P}\delta_{\kappa',\kappa}\delta_{\alpha',\alpha}
 \right]
A^{P}_{n\kappa\alpha} = 0.
\eeq
The non-zero terms appearing in potential matrix element have
$(P',\kappa';P,\kappa)=(+,+;+,+),(+,+;-,-),(+,-;+,-),
(-,+;-,+),(+,-;-,+)$, and $(-,-;-,-)$ and states obtained 
from these by $(P',\kappa') \leftrightarrow (P,\kappa)$.

The $E_{\alpha\kappa}$ are the eigenstates
of the free Dirac Hamiltonian $E_{\alpha\kappa} = p_{\alpha\kappa}$
with momenta $p_{\alpha\kappa}$ fixed by the bag model
boundary condition 
\beq
\label{eqn:bmbc}
\bar{\Phi}^+_{\alpha\kappa}(r)\Phi^+_{\alpha\kappa}(r)\vert_{r=R}=0
\eeq
where $\bar{\Phi}=\Phi^\dagger\gamma^0$ and $r$ is the
bag radius. We consider only 
$\Phi^+_{\alpha\kappa}$ in Eq.(\ref{eqn:bmbc}) since the negative
energy states yield the same momenta $p_{\alpha\kappa}$.

We solve Eq.(\ref{eqn:depb})
in the bag model basis up to some maximum value of the momentum 
$p_{\alpha_m\kappa}$. We include enough states in the
basis so that the longitudinal sum rule, $S_L(\qmag)$
(see Eq.(\ref{eqn:SL}) of the next section) is satisfied.
The value of the radius $R$ is set by the requirement that radii
of all included positive energy eigenstates of $H$ have {\em rms}
radii $\ll R$. $R=15$ fm is used to satisfy this condition.
The negative energy eigenstates have a continuum
of eigenvalues in the limit $R\rightarrow\infty$ and therefore
are $R$-dependent when $R$ is finite. 

The diagonalization of Eq.(\ref{eqn:depb}) is effected for
each negative value of $\kappa$. The spin symmetry admitted
by $H$ allows one to obtain the states with $\kappa>0$ from those with
$\kappa<0$ since the states with $\kappa+\kappa'=-1$ are degenerate.
For each value of $\kappa$, angular integrations in the matrix
elements of the potential in Eq.(\ref{eqn:depb}) are evaluated
explicitly. $H$ is then diagonalized in a truncated basis 
of radial states (spherical Bessel functions appearing in
the bag model states -- see Appendix \ref{sec:app1}) with wave
numbers $\{p_{\alpha\kappa}\}_{\alpha=1}^{\alpha_m}$. In this work
we take $\alpha_m=600$. Thus for each $\kappa$ we have 600 positive
energy and 600 negative energy states in the basis.

The eigenstates of Eq.(\ref{eqn:Deq}) can be written
\beqa
\label{eqn:Psin}
\Psi^{P}_{n\kappa m} &=& \left(\begin{array}{c}
f_{n\kappa}(r)\yjm{\ell}(\hat{\bvec{r}}) \\
ig_{n\kappa}(r)\yjm{\ell\pm 1}(\hat{\bvec{r}}) \end{array}\right)
\eeqa
where the angular part of the wave function is the spin-angle
function, $\yjm{\ell}$, with orbital angular momentum $\ell$ and
spin-$\half$ coupled to total angular momentum $j$ and projection
$m$. The upper (lower) sign in the 
angular part of the lower component of the wave function is
taken when $\kappa$ is negative (positive) with $\kappa=\ell>0$
and $\kappa=-(\ell+1)<0$.
We obtain, from the expansion Eq.(\ref{eqn:exp}) and the explicit
forms of the bag model wave functions Eqs.(\ref{eqn:psip-}) and
(\ref{eqn:psip+}) and their negative energy counterparts, the expressions
for the radial wave functions (for $\kappa<0$)
\beqa
\label{eqn:fkn}
f_{n\kappa}(r) &=& \sum_{\alpha=1}^{\alpha_m}
\left( A^+_{n\kappa\alpha} N_{\alpha\kappa} j_\ell(p_{\alpha\kappa} r)
    -i A^-_{n(-\kappa)\alpha} N_{\alpha(-\kappa)}
       \sqrt{\frac{E_{\alpha(-\kappa)} - m}{E_{\alpha(-\kappa)} + m}}
       j_\ell(p_{\alpha(-\kappa)} r) \right) \\
g_{n\kappa}(r) &=& \sum_{\alpha=1}^{\alpha_m}
\left(-A^+_{n\kappa\alpha} N_{\alpha\kappa}
       \sqrt{\frac{E_{\alpha\kappa} - m}{E_{\alpha\kappa} + m}}
       j_{\ell+1}(p_{\alpha\kappa} r)
    -i A^-_{n(-\kappa)\alpha} N_{\alpha(-\kappa)}
       j_{\ell+1}(p_{\alpha(-\kappa)} r) \right).
\eeqa
Analogous expressions hold for $\kappa>0$.

We note that the ground state has energy $E_0=840$ MeV 
and {\em rms} radius $\langle r^2 \rangle^\half=0.66$ fm
for string tension $\st=1$ GeV/fm. This string tension corresponds
to a characteristic energy scale of $\sigma^{1/4}=440$ MeV. 
On the basis of dimensional analysis we find that 
$E_0\propto \sigma^{\frac{1}{4}}$.

\section{Electromagnetic structure functions}
\label{sec:sf}
The inclusive cross section for a spin-$\half$ electromagnetic probe 
scattering from a hadronic target may be expressed 
to first order in the fine structure constant $\alpha$ as
\beqa
\label{eqn:xs12}
\frac{d^2\sigma}{dE'd\Omega}
&=& \sigma_M \left( W_2 + 2 W_1 \tan^2(\theta/2) \right) \\
&=& \sigma_M \frac{Q^2}{\qmag^2}
    \left[W_L+W_T\left(1+2\frac{\qmag^2}{Q^2}\tan^2(\theta/2) \right)\right].
\eeqa
Here $\sigma_M$ is the Mott cross section
\beq
\label{eqn:mott}
\sigma_M = \frac{\alpha_q^2 \cos^2(\theta/2)}{4E^2\sin^4(\theta/2)},
\eeq
with $\theta$ the probe's scattering angle, $E(E')$ its initial
(final) energy and $\alpha_q=Q_q\alpha$, $Q_q$ is the charge on the
hadronic target's constituent in units of the proton's charge and
$\alpha$ is the fine structure constant.
We have neglected the mass of the electromagnetic probe
in the above. The inclusive longitudinal, $W_L$ and transverse,
$W_T$ structure functions are related to the functions $W_{1,2}$ by
\beqa
\label{eqn:WT}
W_T &=& W_1 \\
\label{eqn:WL}
W_L &=& -W_1 + \frac{\qmag^2}{Q^2} W_2.
\eeqa
We have assumed that $q^\mu=(\nu,0,0,\qmag)$ and 
defined $Q^2=-q^2=\qmag^2 - \nu^2$.

\begin{figure}
\includegraphics[ width=275pt, keepaspectratio, angle=0, clip ]{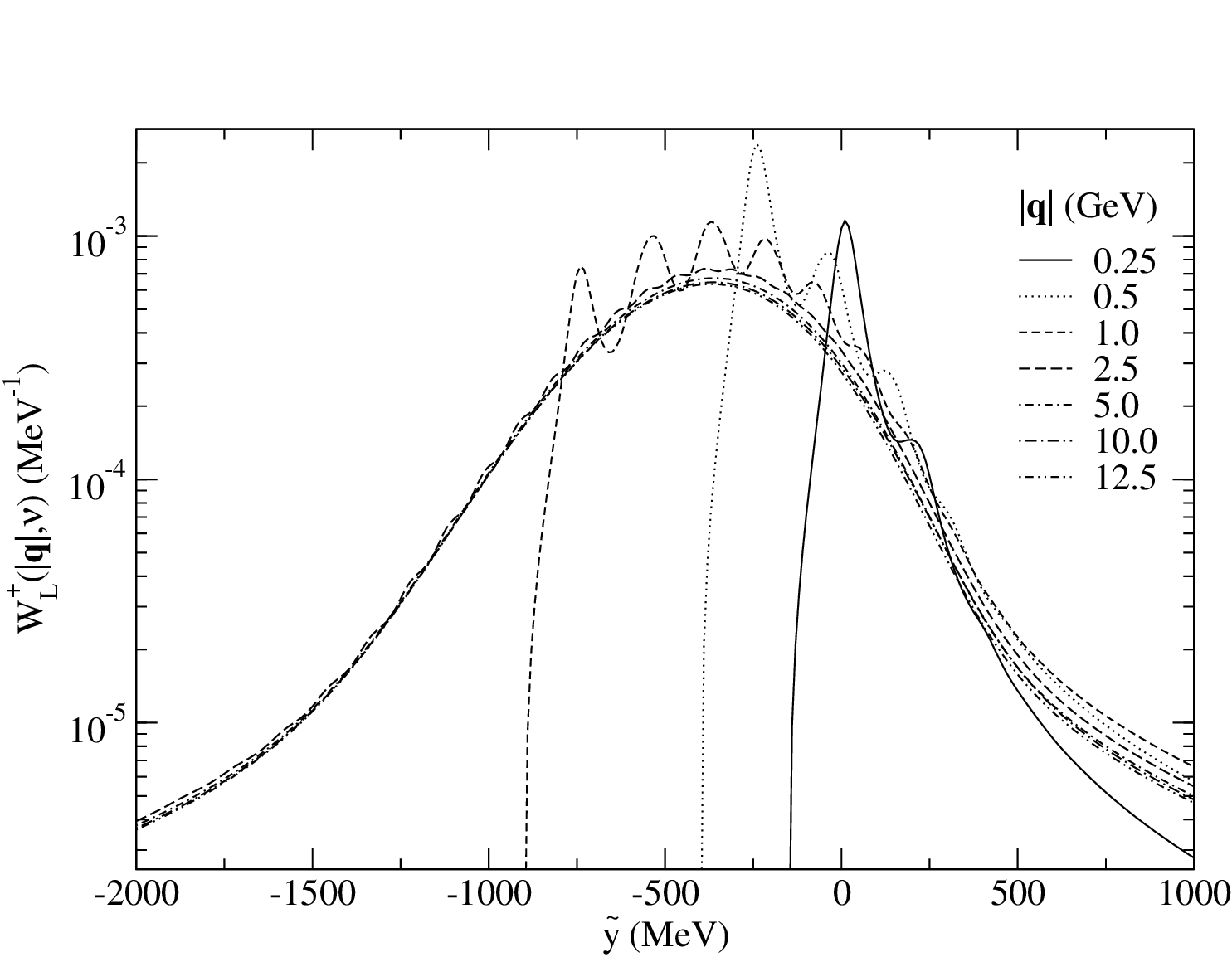}
\caption{\label{fig:WLvq} The longitudinal structure function 
$W^+_L(\qmag,\nu)$ vs. $\ysc=\nu-\qmag$ for various values of the 
three-momentum transfer.}
\end{figure}

The inclusive structure functions are determined from the
matrix elements of the hadronic current on the target eigenstates 
and their energy eigenvalues as
\beqa
\label{eqn:WL+}
W^+_L(\qmag,\nu) &=& \half \sum_{m,I}
\left|\bra{I}\Lambda_+ e^{i\bvec{q}\cdot\rvec}\ket{0,m}\right|^2
\delta(E_I-E_0-\nu), \\
\label{eqn:WT+}
W^+_T(\qmag,\nu) &=& \half \sum_{m,I}
\left|\bra{I}\Lambda_+\alpha_+ e^{i\bvec{q}\cdot\rvec}\ket{0,m}\right|^2
\delta(E_I-E_0-\nu)
\eeqa
where $\ket{0,m}$ is the ground state with spin projection
$m=\pm\half$, $\alpha_+=\gamma^0\gamma_+$,
$\gamma_+=\frac{1}{\sqrt{2}}(\gamma^1+i\gamma^2)$,
and the $\frac{1}{2}\sum_{m,I}$ averages over initial spins
and sums over final states, $\ket{I}$.
We assume that the negative energy states are occupied and 
thus excluded from the sum over final states by Pauli exclusion. 
The $+$ superscript on the structure functions indicates this
exclusion. The projection operators on positive and negative
energy states are defined by
\beqa
\label{eqn:Lam+}
\Lambda_+ &=& \sum_{I,E_I\geq E_0} \ket{I}\bra{I} \\
\label{eqn:Lam-}
\Lambda_- &=& \sum_{I,E_I < E_0} \ket{I}\bra{I},
\eeqa
respectively. The states
$\ket{I}$ are orthonormal $\bb{I}|J\kb=\delta_{IJ}$.
In the interest of subsequent discussion we also define
$W^-_{L,T}$ with $\Lambda_-$ in Eqs.(\ref{eqn:WL+}) and
(\ref{eqn:WT+}) instead of $\Lambda_+$.

The longitudinal and transverse structure functions $W^+_{L,T}$
are related to the response of the electromagnetic probe to 
the charge and spin fluctuations in the target, respectively. They are 
viewed as the response of the light, positive energy valence quark 
in a heavy-light meson, such as $\bar{t}u$, as it
is excited into a positive energy final state within the potential well.
The longitudinal function $W^+_L$ is associated with a spin-independent
coupling of the probe to the target and thus the spin of the struck
constituent is unchanged in contributions to it.
The transverse function $W^+_T$ is associated with a spin-flip of the 
struck constituent and couples the upper and lower components of 
its wave function since
\beqa
\label{eqn:gam+}
\gamma_+=\left( \begin{array}{cc} 0 & \sqrt{2}\sigma_+ \\
 -\sqrt{2}\sigma_+ & 0 \end{array} \right),
\eeqa
where $\sigma_+$ is the raising operator on Pauli spinors.

We calculate the structure functions in a large basis of
eigenstates of the Dirac Hamiltonian [Eq.(\ref{eqn:Deq})]
including values of $\kappa$ in the range $-100 \leq \kappa < 0$
and $0<\kappa\leq 100$ in the sum over $I$ in Eqs.(\ref{eqn:WL+},
\ref{eqn:WT+}).  We take $400$ positive (negative) energy radial 
states each for each value of $\kappa$ for the calculation of 
$W^+(W^-)$. We ensure that this basis
supports all the states accessed by scattering up to $\qmag=12.5$
GeV to a level of $0.3\%$ as measured by the static
structure functions, $S_{L,T}(\qmag)$ (as shown below).

In order to obtain the structure functions as smooth curves
we fold them with a Breit-Wigner function letting
\beq
\label{eqn:br-wig}
\delta(E_I-E_0-\nu) \rightarrow
\frac{\Gamma(\nu)}{2\pi}\frac{1}{(E_I-E_0-\nu)^2 + \Gamma^2(\nu)/4},
\eeq
with level width $\Gamma(\nu)$. We use the parameterization for 
$\Gamma(\nu)$ given in Eq.(8) of Ref.\cite{Paris01} with 
$\Gamma_0=100$ MeV. The effects of hadronization of the excited
states are taken into account qualitatively by this prescription.

Figure (\ref{fig:WLvq}) shows $W^+_L(\qmag,\nu)$ for various 
values of $\qmag$ plotted as a function of the variable
$\ysc=\nu-\qmag$. We note some features of the $W^+_L$. 
The curves at high $\qmag\gtrsim 2.5$ GeV, much larger than the
characteristic energy scale of $\sigma^{1/4}=440$ MeV, exhibit scaling in 
the variable $\ysc$. There is non-zero strength in the timelike 
region where $\ysc>0$ corresponding to the fact that the confined 
constituent is {\em bound} in the target. The strength in the 
timelike region for $\qmag=12.5$ GeV is 7.3\% and 11.5\% for 
$\Gamma_0=0$ and 100 MeV, respectively. 
At $\qmag\lesssim 2.5$ GeV the curves exhibit 
resonances for $\ysc\sim -\qmag$ and show Bloom-Gilman 
duality.\cite{Bloom,Paris01} The quality
of scaling is fair though scaling violations are significant 
in the timelike region having $\ysc>0$ in the present
$\qmag$ range. In order to understand these violations
we will discuss sum rules.

The full strength of the ground state is distributed among the
the positive and negative energy eigenstates by the current
$\gamma^\mu e^{i\bvec{q}\cdot\rvec}$. Therefore both $W^+_L$
and $W^-_L$ must be taken into account to support the full
strength of the ground state. If we write
\beq
\label{eqn:WLpn}
W_L(\qmag,\nu) = W^+_L(\qmag,\nu) + W^-_L(\qmag,\nu)
\eeq
we obtain a function which satisfies the sum rule
\beq
\label{eqn:SL}
S_L(\qmag) = \int_{-\infty}^\infty d\nu W_L(\qmag,\nu) = 1,
\eeq
where $S_L(\qmag)$ is the longitudinal static structure function.
Here we have integrated over all values of energy transfer $\nu$
including the region $\nu<0$ and used completeness
\beq
\label{eqn:cmplt}
\sum_I \ket{I}\bra{I} = 1.
\eeq
(In fact, in our numerical study, $S_L(\qmag)$ has the value of 
unity to $0.3\%$ or better for all considered $\qmag$.)

The total strength of the $W^+_L(\qmag,\nu)$
\beq
\label{eqn:SL+}
S_L^+(\qmag) = \int_0^\infty W^+_L(\qmag,\nu) < 1
\eeq
since we have neglected the contribution of the negative
energy states. As $\qmag$ increases, the
strength $S^+_L(\qmag)$ due to the positive energy states decreases.
It can be shown (see Appendix \ref{sec:app2}) that
\beq
\label{eqn:SL+lim}
\lim_{\qmag\rightarrow\infty}S_L^+(\qmag) = \half.
\eeq
At $\qmag=12.5$ GeV the $S^+_L(\qmag)=0.52$, independent of 
$\Gamma_0$.  The strength missing from $S_L^+(\qmag)$ is taken 
up by the negative energy states
\beq
\label{eqn:SLn}
S^-_L(\qmag) = \int_{-\infty}^0 d\nu W^-_L(\qmag,\nu)
\eeq
to maintain the sum rule Eq.(\ref{eqn:SL}). The scaling violations
in $W^+_L$ are due to the sharing of the strength of the ground
state between the positive and negative energy eigenstates of the
potential. 

\begin{figure}
\includegraphics[ width=275pt, keepaspectratio, angle=0, clip ]{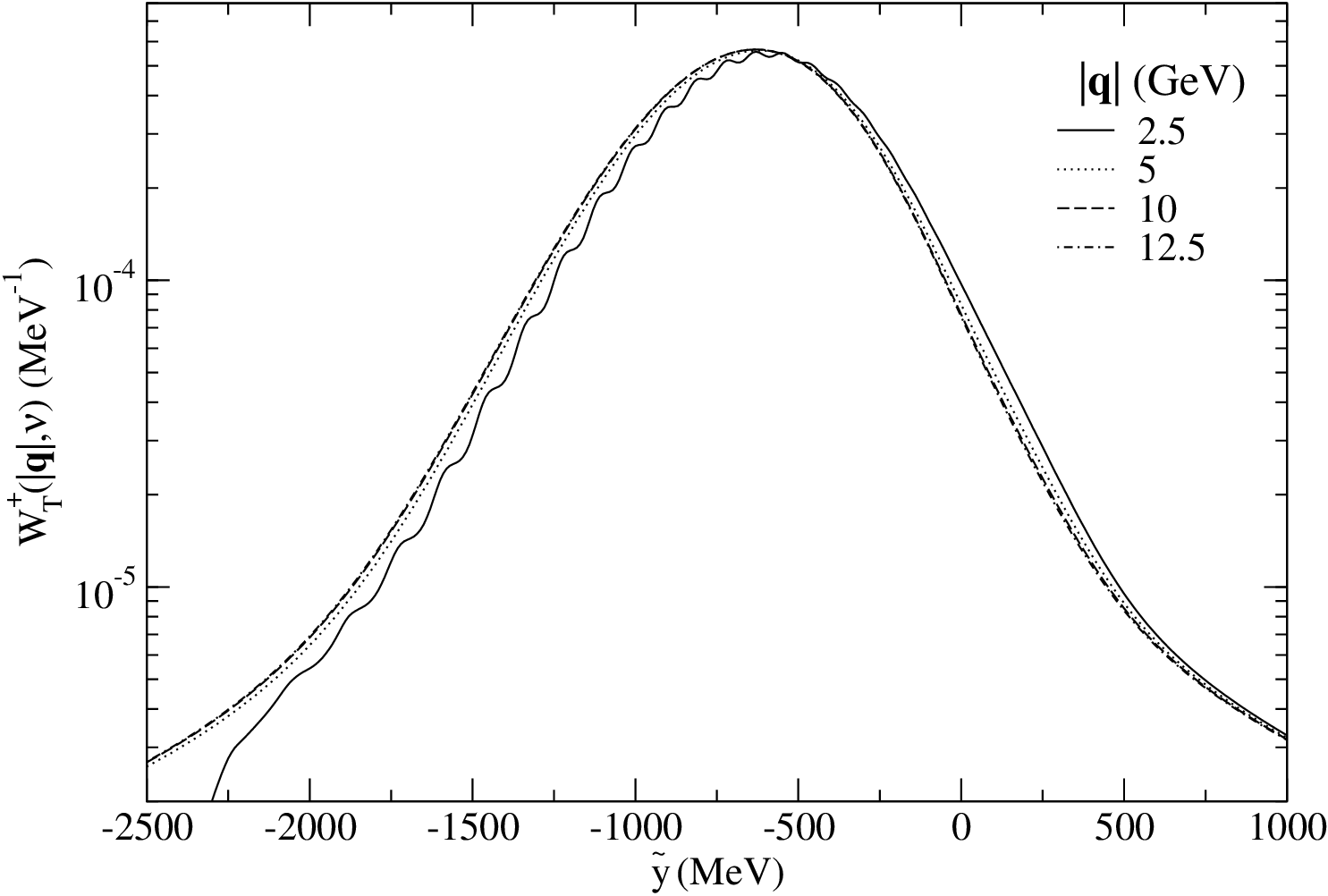}
\caption{\label{fig:WT+} The transverse structure function 
$W^+_T(\qmag,\nu)$
vs. $\ysc=\nu-\qmag$ for four values of $\qmag=2.5, 5, 10, 12.5$ GeV.}
\end{figure}

We expect similar scaling violations in the case of the transverse 
structure function $W_T^+(\qmag,\nu)$ which we have shown in 
Fig.(\ref{fig:WT+}). The strength in the timelike region for $\qmag=12.5$
GeV is 1.4\% and 4.5\% for $\Gamma_0=0,100$ MeV, respectively.

The transverse structure function $W_T(\qmag,\nu)$, defined
analogously to Eq.(\ref{eqn:WLpn}), satisfies the sum rule
\beqa
\label{eqn:ST}
S_T(\qmag) &=& \int_{-\infty}^\infty d\nu W_T(\qmag,\nu) \\
&=& \half\sum_m \bra{0,m}(1-\sigma_z)\ket{0,m} \nonumber \\
\label{eqn:STres}
&=& \frac{5}{6} + \frac{1}{6} = 1,
\eeqa
where the first and second terms in Eq.(\ref{eqn:STres})
correspond to $m=-\half$ and $m=+\half$, respectively.
The different strengths in $m=\pm\half$ can be understood
by analyzing the ground state wave function. 
The matrix operator $\alpha_+$ in Eq.(\ref{eqn:WT+})
corresponds to a virtual photon with positive 
helicity and therefore can only scatter from
components of the wave function which have Pauli-spin down.
The ground state wave function is $s$-wave in the upper 
component and the photon can't scatter from it
when $m=+\half$. It only scatters from the spin down part 
of the $p$-wave lower component resulting in less strength 
for the $m=+\half$ state than in the $m=-\half$ state.

In deriving the result Eq.(\ref{eqn:STres}) we used the facts that
\beqa
\label{eqn:gsprops}
\int_0^\infty dr\ r^2 f_0^2(r) = \frac{3}{4} \\
\int_0^\infty dr\ r^2 g_0^2(r) = \frac{1}{4}
\eeqa
where $rf_0(r)$ and $rg_0(r)$ are the upper and lower radial 
wave functions for the ground state. These follow directly 
from the Dirac equation for a massless constituent independent 
of the string tension $\st$ as illustrated in Appendix \ref{sec:app3}.
Additionally it can be shown that
\beq
\label{eqn:virial}
E_0 = 2\gsa{V},
\eeq
where $\gsa{O}\equiv\int d^3r \Psi^\dagger_0(\rvec) O \Psi_0(\rvec)$.

Up to this point we have calculated the structure functions
$W^+_{L,T}$ for a confined Dirac valence particle. These functions
concern the scattering of the valence particle in the ground state
to a positive energy excited state. The contribution of the negative
energy states in the sum over intermediate states $I$ in 
Eq.(\ref{eqn:WL+}) has been ignored since these states are assumed 
to be filled. 

\begin{figure}
\includegraphics[ width=275pt, keepaspectratio, angle=0, clip ]{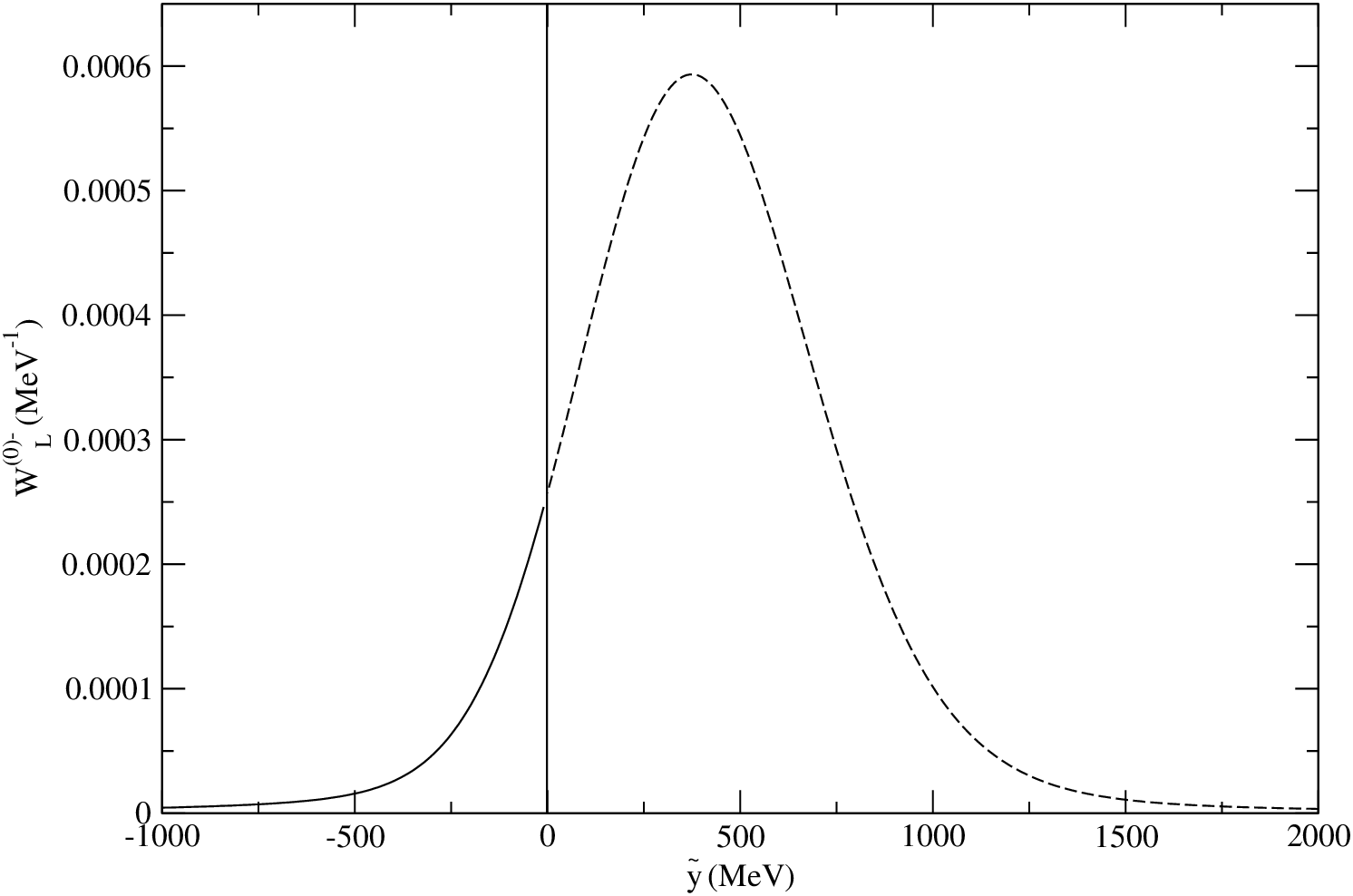}
\caption{\label{fig:WL-} The vacuum contribution to the longitudinal 
structure function $W^{(0)-}_L(\qmag,\nu)$ for scattering from a negative 
energy state into the ground state vs. $\ysc=\nu-\qmag$ for $\qmag=10$ GeV.
Only the solid curve in the region $\ysc<0$ is physical. The continuation
(dashed curve) into the region $\ysc>0$ has no meaning.}
\end{figure}

We now consider the response of the vacuum in the presence of the
infinitely massive color source. The response of the vacuum 
corresponds to scattering from a negative energy 
state into a positive energy state. Consider the contribution
to the vacuum response when the particle is initially in a negative
energy state $I$ and scatters into the unoccupied ground state:
\beq
\label{eqn:WL-}
W^{(0)-}_L(\qmag,\nu) = \half \sum_{m,I}
|\bra{0,m}e^{-i\bvec{q}\cdot\rvec}\Lambda_-\ket{I}|^2
\delta(E_I-E_0+\nu).
\eeq
Here the superscript $(0)-$ indicates that we are scattering from
a negative energy state into the ground state. We have taken 
the energy transfer $\nu\rightarrow -\nu$ since a 
negative energy state is equivalent to a positive energy antiparticle.
This contribution is shown Fig.(\ref{fig:WL-}) as a function of $\ysc$ 
for $\qmag=10$ GeV. We noted above that the negative energy states
depend on the cavity radius, $R$, since they are not bound by the
potential Eq.(\ref{eqn:potdef}). However
$W^{(0)-}_L$ is only weakly dependent on the cavity radius since
the matrix element receives contributions for $r\lesssim 1$ fm,
the radial extent of the ground state. At these radii the negative
energy states are not sensitive to the boundary condition. This
was verified for 
$R=10$ and 15 fm. We interpret $W^{(0)-}_L$  as the contribution to the
vacuum response in the spacelike $\ysc<0$ region. It is a consequence
of the fact that interactions move some of the strength of the 
vacuum response from the timelike region into the spacelike region.
We cannot reliably calculate the response in the timelike region since 
confinement acts at all distances in our model. Additional contributions 
to the vacuum response in the spacelike region correspond to scattering
from a negative energy state into a positive energy state $J$ and
are denoted $W^{(J)-}_L$. These contributions are smaller than the
$W^{(0)-}_L$ since the first excited state has energy $E_{J=1}=1100$ MeV,
$260$ MeV above the ground state.

\section{Plane wave impulse approximation}
\label{sec:PWIA}
We may compare the exact structure functions calculated in the
previous section with those obtained in PWIA. We consider only
the longitudinal structure function in this section. In PWIA, 
one assumes that the state of the struck constituent may be 
described as a plane wave with energy
\beq
\label{eqn:pwen}
E_{\bvec{k}+\bvec{q}} = |\bvec{k}+\bvec{q}| + \gsa{V},
\eeq
for a massless constituent. The expectation value of the potential 
[Eq.(\ref{eqn:potdef})]
in the ground state, $\gsa{V}$ is required to give the correct
value for the energy weighted sum rule
\beq
\label{eqn:S1Lq}
S^{(1)+}_L(\qmag)=\int_0^\infty d\nu\,\nu W^+_L(\qmag,\nu).
\eeq
We may calculate this quantity analytically in the limit
$\qmag\rightarrow\infty$ using the technique in Appendix \ref{sec:app2}
to obtain 
\beq
\label{eqn:S1Lqlim}
\lim_{\qmag\rightarrow\infty}S^{(1)+}_L(\qmag) = \frac{\qmag}{2}.
\eeq

The PWIA neglects interactions 
of the struck constituent in the final state. The expression 
Eq.(\ref{eqn:WL+}), for example, then simplifies to
\beq
\label{eqn:WL+PW}
W^+_{L,PWIA}(\qmag,\nu) = \half\sum_{s,m}\int\frac{d^3k}{(2\pi)^3}
\left|\bra{u_{\bvec{k}+\bvec{q},s}}e^{i\bvec{q}\cdot\rvec}\ket{0,m}\right|^2
\delta(|\bvec{k}+\bvec{q}|+\gsa{V}-E_0-\nu),
\eeq
where $\ket{u_{\bvec{k}+\bvec{q},s}}$ is the positive energy
free-particle Dirac spinor with spin $s$. An analogous expression
holds for the $W^+_{T,PWIA}$.

\begin{figure}
\includegraphics[ width=275pt, keepaspectratio, angle=0 ]{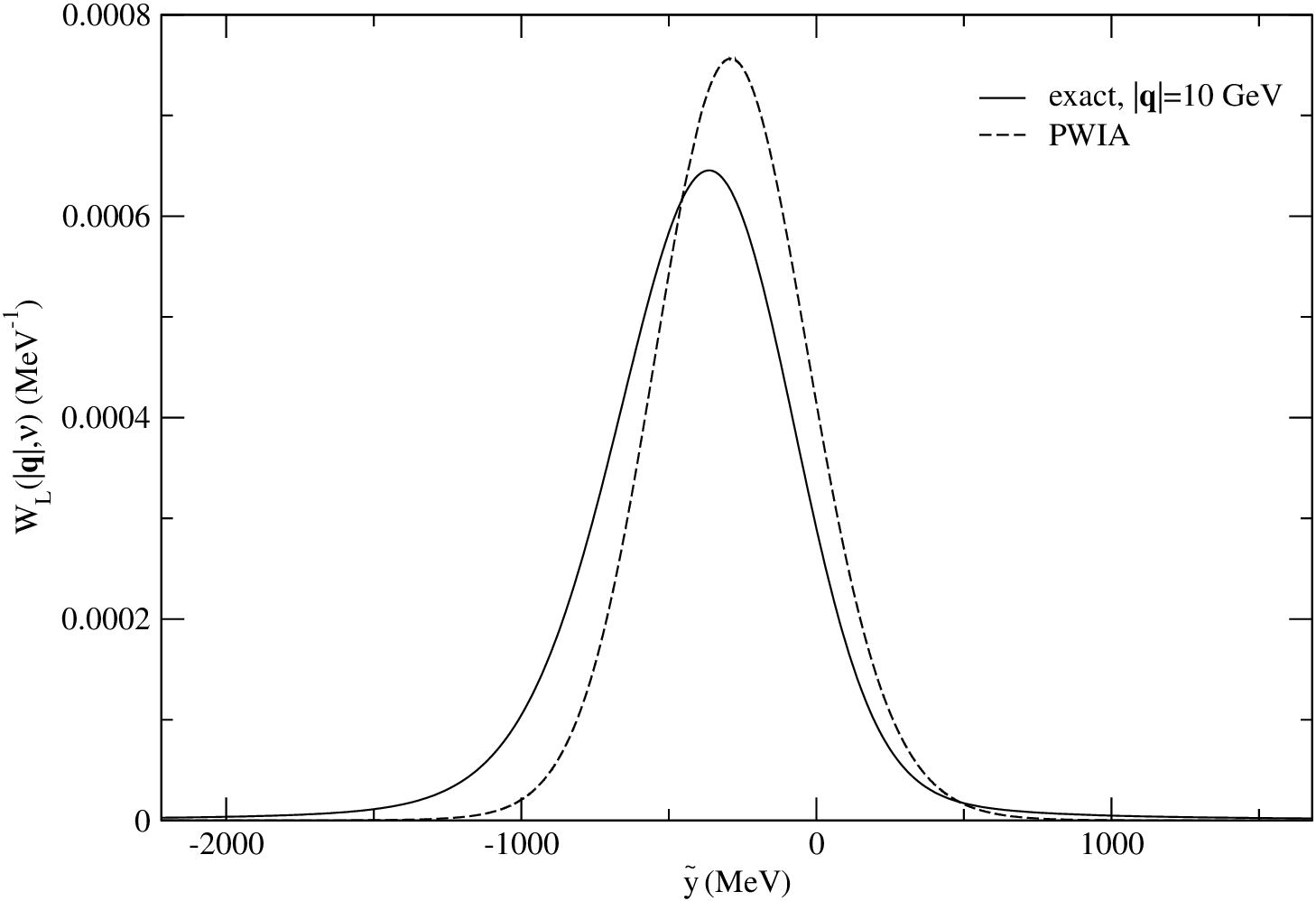}
\caption{\label{fig:wlpwvsx} The longitudinal structure function 
$W^+_L(\qmag,\nu)$ vs. $\ysc=\nu-\qmag$ in PWIA and the exact result
for $\qmag=10$ GeV.}
\end{figure}

Evaluating the matrix elements in the above equation gives
\beqa
\label{eqn:WL+PWme}
W^+_{L,PWIA}(\ysc) &=& \half\int\frac{d^3k}{4\pi}
\left[\tilde{f}^2_0(k)+\tilde{g}^2_0(k)
-2\tilde{f}_0(k)\tilde{g}_0(k)\frac{k_z}{k}\right] \nonumber \\
&\times&\delta(k_z-\ysc+\gsa{V}-E_0),
\eeqa
where we have neglected terms $\mathcal{O}(\frac{1}{\qmag})$ and
$\tilde{f}_0(k)$, $\tilde{g}_0(k)$ are the Fourier-Bessel transforms
of the radial wave functions appearing in the Fourier transform of 
the ground state:
\beq
\label{eqn:gsft}
\tilde{\Psi}_{0,m}(\bvec{k}) = (2\pi)^\frac{3}{2} \left(
\begin{array}{c} \tilde{f}_0(k) \ysa{\half}{m}{0}(\hat{\bvec{k}}) \\
\tilde{g}_0(k) \ysa{\half}{m}{1}(\hat{\bvec{k}}) \end{array}\right).
\eeq
The $\yjm{\ell}$ are the spin-angle functions.  Figure 
\ref{fig:wlpwvsx} shows the PWIA approximation for the longitudinal
structure function and the exact result.

The argument of the $\delta$ function depends only on the variable 
$\ysc$ and therefore the PWIA response also exhibits scaling behavior. 
It has scaling violations which are $\mathcal{O}(\frac{1}{\qmag})$ 
for $\qmag$ finite. The $W^+_{L,PWIA}$ satisfies the sum rules
Eq.(\ref{eqn:SL+lim}) and Eq.(\ref{eqn:S1Lq}) but has a different
shape than the exact curve. It has more response in the timelike region
than the exact result with 14.7\% of the total strength in the
timelike region compared with 11.5\% for the exact result. This
is a consequence of the interference of the upper and lower
components of the ground state wave function in 
the third term in square brackets of Eq.(\ref{eqn:WL+PWme}).
Neglecting this term, the $W^+_{L,PWIA}$ would peak at 
$\ysc=-\frac{E_0}{2}$, nearly coincident with the peak of the
exact curve. The interference term shifts strength into the timelike
region. In a subsequent work we shall explore the consequences of 
such effects on spin-dependent observables such as the helicity 
structure function.\cite{PPS03} We note finally that the PWIA and
exact curves in Fig.(\ref{fig:wlpwvsx}) have different widths.
The larger width of the exact curve is due to FSI of the struck
constituent with the potential, $V$. As in the case of the 
semirelativistic model \cite{Paris01}, FSI effects persist in
the limit $\qmag\rightarrow\infty$. We note that Brodsky and
collaborators have shown that FSI affect the interpretation of 
parton distribution functions \cite{Brodsky02} and can contribute
to large single spin asymmetries in semi-inclusive DIS processes
\cite{Brodsky03}.

\begin{figure}
\includegraphics[ width=275pt, keepaspectratio, angle=0, clip ]{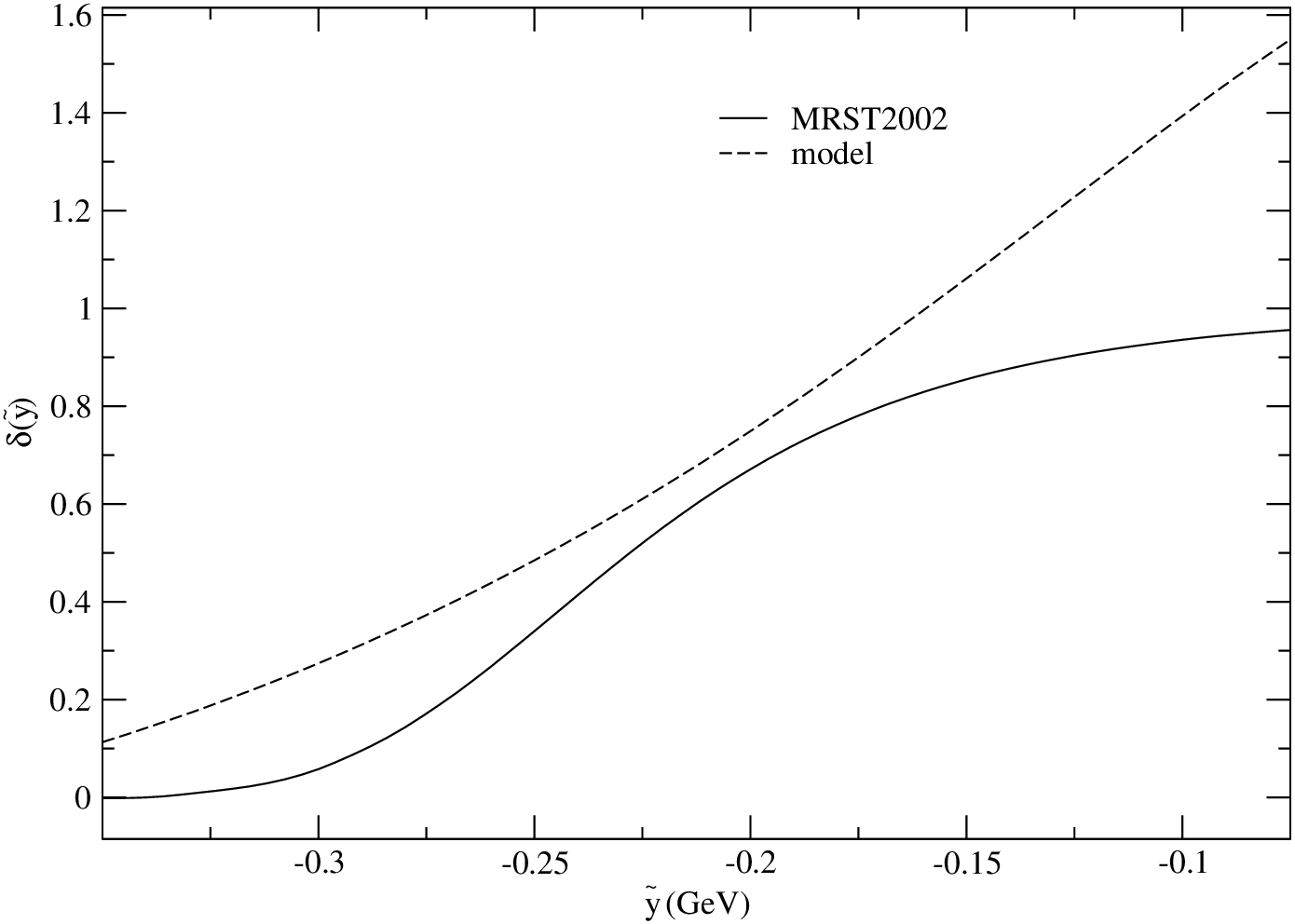}
\caption{\label{fig:cgdev} The deviation from Callan-Gross relation
in terms of the function $\delta$ of Eq.(\ref{eqn:CGdev}) for
$\qmag=10$ GeV. The dashed curve is the scaled model calculation (see text).
The solid curve is the MRST2002 {\em pdf} fit to data.}
\end{figure}

\section{Conclusion}
\label{sec:concl}
We have calculated exactly the eigenstates of a massless Dirac 
particle confined in a linear potential for excitation energies up to
$\sim 10$ GeV. The unpolarized valence quark
structure functions of the particle
in the ground state scattering into a positive energy excited 
state were determined by coupling a spin-$\half$
electromagnetic probe to the 
conserved vector current of particles in the potential well. 
The structure functions exhibit $\ysc$ or Nachtmann
scaling, a generalization 
of Bjorken scaling. The exact response takes account of interactions
of the confined particle both before and after interaction with
the electromagnetic probe and therefore $\ysc$ or Bjorken scaling
is not a consequence of treating the constituent of the target as
a free particle. In contrast to Refs.\cite{Pace98,Jesch02} who find that
FSI have no effect on the structure functions, and therefore
also conclude that Bjorken scaling applies to systems of interacting
constituents, the present model does not reproduce the
parton model results when FSI are taken into account.

Another example of the role interactions have in determining the
structure function of confined Dirac particles can be seen if we
consider the function
\beq
\label{eqn:CGdev}
\delta(\ysc,\qmag) = \frac{F_2(\ysc,\qmag)-2xF_1(\ysc,\qmag)}
{F_2(\ysc,\qmag)+2xF_1(\ysc,\qmag)}
\eeq
where $F_1 = M W_1$ and $F_2=\nu W_2$,
$M$ is the mass of the target.  At leading order in the quark-parton model
one obtains $\delta(\ysc,\qmag)=0$ (the Callan-Gross relation \cite{CG})
for all $\ysc$ and $\qmag$. In Fig.(\ref{fig:cgdev}) we have plotted,
for $\qmag=10$ GeV,
the MRST2002 \cite{MRST02} fit (solid curve) to deep inelastic data 
using the next-to leading order prediction \cite{Cooper88}
\beq
\label{eqn:FL}
F_L(x,Q^2) = \frac{2.5}{5.9}\frac{4\pi}{\alpha_s(Q^2)}xg(2.5x,Q^2),
\eeq
where $0<x<0.4$ and $g(x,Q^2)$ is the prediction for the gluon. It is 
compared to the present model calculation (dashed curve) which has 
been rescaled in order to obtain agreement with the first moment 
of the {\em up} valence quark distribution for the MRST {\em pdf}
assuming that the response of the proton is due to a single 
{\em up} valence quark. The non-perturbative effects of interactions
could be the origin of the difference in the two curves in 
Fig.(\ref{fig:cgdev}).

In future work we will apply the present model to the study
of spin-dependent effects such as the helicity and transversity
distribution of the constituents of the target. We expect that
these observables depend strongly upon the relativistic treatment
of the constituent as a Dirac particle. The comparison of the
present model with experimental data requires one to include 
radiative gluon effects to obtain the $Q^2$ evolution of the
structure functions. This problem is currently under study.

\appendix

\section{Bag model basis states}
\label{sec:app1}
The free Dirac Hamiltonian $H_0$ is
\beqa
\label{eqn:Deq0}
\left[\bfalp\cdot\bvec{p}+\beta m\right] \Phi^{P}_{\alpha\kappa m}
&=&\mbox{sgn}P\sqrt{(p_{\alpha\kappa})^2+m^2}
\, \Phi^{P}_{\alpha\kappa m} \nonumber \\
&=&\mbox{sgn}P E_{\alpha\kappa} \Phi^{P}_{\alpha\kappa m},
\eeqa
for a particle of mass $m$ (not to be confused with the total spin
projection which appears as the subscript $m$ in the wave function 
$\Phi^P_{\alpha\kappa m}$).
The complete set of states includes states of positive $\eka>0$ energy
states, `particles' and negative $\eka<0$ energy states, `antiparticles.'
The bag model states are solutions of Eq.(\ref{eqn:Deq0}) within
a cavity of radius $R$.  These particle states may be written as
\beqa
\label{eqn:psip-}
\Phi_{\alpha\kappa m}^{+}(\rvec) = N_{\alpha\kappa} \left( \begin{array}{c}
j_\ell(p_{\alpha\kappa}r)\yjm{\ell} \\
-i\sqrt{\frac{\eka-m}{\eka+m}}
j_{\ell+1}(p_{\alpha\kappa}r)\yjm{\ell+1} \end{array}\right), \kappa<0 \\
\label{eqn:psip+}
\Phi_{\alpha\kappa m}^{+}(\rvec) = N_{\alpha\kappa} \left( \begin{array}{c}
j_\ell(p_{\alpha\kappa}r)\yjm{\ell} \\
i\sqrt{\frac{\eka-m}{\eka+m}}
j_{\ell-1}(p_{\alpha\kappa}r)\yjm{\ell-1} \end{array}\right), \kappa>0.
\eeqa
Here the normalization $N_{\alpha\kappa}$ is fixed by the condition
$\int d^3r \Phi^{+\dagger}_{\alpha\kappa m}(\rvec)
  \Phi^{+}_{\alpha\kappa m}(\rvec) = 1$ to give
\beq
\label{eqn:nrmk}
N_{\alpha\kappa}
= \left\{ R^3 \left[ \frac{2\ekaq}{(p_{\alpha\kappa})^2}
+ \frac{1}{(p_{\alpha\kappa})^2 R}(m + 2\eka\kappa) \right]
j^2_{\pm(\kappa+1)}(p_{\alpha\kappa} R) \right\}^{-\frac{1}{2}}.
\eeq
The upper (lower) sign corresponds to $\kappa$ positive (negative).

The antiparticle (negative energy) states are obtained from
Eqs.(\ref{eqn:psip-}) and (\ref{eqn:psip+}) by interchanging
upper and lower components and multiplying one of them by $-1$.
With this prescription the sign of $\kappa$ for the negative
energy state is opposite to the sign of $\kappa$ of the positive
energy state from which it was obtained.

As mentioned in the text, we determine the 
allowed $p_{\kappa\alpha}$ by subjecting the 
$\Phi_{\alpha\kappa m}$ to the bag model boundary condition at
$r=R$
\beq
\label{eqn:bmbcapp}
\left.\overline{\Phi}\Phi \right|_{r=R} = 0,
\eeq
where $\overline{\Phi} = \Phi^\dagger \beta$.
Substitution of the wave functions in Eqs.(\ref{eqn:psip-})
and (\ref{eqn:psip+}) into this expression results in the
following transcendental equations; for $\kappa=-(\ell+1)<0$
\beq
\label{eqn:bc-}
j_\ell(p_{\alpha\kappa} R) =
\sqrt{\frac{\eka-m}{\eka+m}} j_{\ell+1}(p_{\alpha\kappa}R),
\eeq
and for $\kappa=\ell>0$
\beq
\label{eqn:bc+}
j_\ell(p_{\alpha\kappa} R) =
-\sqrt{\frac{\eka-m}{\eka+m}} j_{\ell-1}(p_{\alpha\kappa}R).
\eeq
The subscript $\alpha=1,\ldots,\infty$ indexes the solutions
of the transcendental equations. The number of nodes in the 
upper component radial function is $(\alpha-1)$.

\section{$S^+_L(\qmag)$ in the limit $\qmag\rightarrow\infty$}
\label{sec:app2}
The expression for the $S_L^+(\qmag)$ is given by
\beqa
\label{eqn:SLq1}
S_L^+(\qmag) &=& \int_0^\infty d\nu W^+_L(\qmag) \\
\label{eqn:SLq2}
&=& \half\sum_m\bra{0,m}e^{-i\bvec{q}\cdot\rvec}\Lambda_+
e^{i\bvec{q}\cdot\rvec}\ket{0,m}.
\eeqa
We may evaluate this expression analytically in the limit
$\qmag\rightarrow\infty$ by recognizing that in this limit 
the intermediate states $\ket{I}$ in 
Eq.(\ref{eqn:SLq2}) may be taken to be free-particle solutions,
$\ket{u_{\bvec{k}+\bvec{q},s}}$
of the Dirac equation since their overlap with the ground state 
is dominated by small radii $\lesssim 1$ fm where the exact
wave functions resemble free-particle wave functions. Thus
we replace
\beq
\label{eqn:approxcmplt}
e^{-i\bvec{q}\cdot\rvec}\Lambda_+ e^{i\bvec{q}\cdot\rvec}
\rightarrow
e^{-i\bvec{q}\cdot\rvec}
\sum_s\int\frac{d^3k}{(2\pi)^3} \ket{u_{\bvec{k}+\bvec{q}},s}
\bra{u_{\bvec{k}+\bvec{q}},s} e^{i\bvec{q}\cdot\rvec},
\eeq
valid in the limit $\qmag\rightarrow\infty$
and obtain upon substitution into Eq.(\ref{eqn:SLq2})
\beqa
\label{eqn:SLq3}
S^+_L(\qmag) &=& \half\sum_m \int\frac{d^3k}{(2\pi)^3}
\Psi^\dagger_{0,m}(\bvec{k})
\sum_s u_{\bvec{k}+\bvec{q},s}u^\dagger_{\bvec{k}+\bvec{q},s}
\Psi_{0,m}(\bvec{k}).
\eeqa
The projection operator for free-particle solutions to the Dirac
equation, expressed to $\oer{}$, is
\beq
\label{eqn:Lfree}
\sum_s u_{\bvec{k}+\bvec{q},s}u^\dagger_{\bvec{k}+\bvec{q},s}
= \half\left(1+\alpha_z+\frac{\bfalp\cdot\bvec{k}_\perp}{\qmag}\right),
\eeq
where $\bvec{q}=\qmag\hat{\bvec{z}}$ and $\bvec{k}_\perp=(k_x,k_y)$.
Substitution of Eq.(\ref{eqn:Lfree}) into Eq.(\ref{eqn:SLq3}) gives
the result Eq.(\ref{eqn:SL+lim}).

\section{Ground state normalizations}
\label{sec:app3}
The Dirac equation for the ground state gives
\beqa
\label{eqn:Deqgsf}
f_0'(r) &=& E_0 g_0(r) \\
\label{eqn:Deqgsg}
-g_0'(r)-\frac{2}{r}g_0(r) &=& (E_0-\st r) f_0(r).
\eeqa
for the upper and lower radial wave functions.
Multiplying Eq.(\ref{eqn:Deqgsf}) by $g_0(r)$ and Eq.(\ref{eqn:Deqgsg})
by $f_0(r)$, summing and integrating gives
\beq
\label{eqn:gsintrel}
E_0\int_0^\infty dr r^2 (f_0^2(r)-g_0^2(r))
= \st \int_0^\infty dr r^2 f_0^2(r) r.
\eeq
The r.h.s.\ of the above is the expectation value of the
potential $V(r)=\st r(1+\beta)$ in the ground state since it
doesn't couple to the lower components of the wave function.
Substitute Eq.(\ref{eqn:virial}) into Eq.(\ref{eqn:gsintrel}) 
to obtain
\beq
\label{eqn:radfnnorm}
\int_0^\infty dr r^2 (f_0^2(r)-g_0^2(r)) = \half.
\eeq
The normalization condition is:
\beq
\label{eqn:norm}
\int_0^\infty dr r^2 (f_0^2(r)+g_0^2(r)) = 1.
\eeq
Adding and subtracting these equations gives the result
Eqs.(\ref{eqn:gsprops}), independent of the string tension, $\st$.

\acknowledgments
The author wishes to thank the {\em Schweizerische Nationalfonds}
for support and Omar Benhar, Joseph Carlson, Vijay Pandharipande,
and Ingo Sick for many useful discussions. This work has been 
supported by U.S.\ Department of Energy under contract W-7405-ENG-36.
 
\bibliography{drcrp}

\end{document}